\documentstyle[11pt]{article}
\newcommand{\pul}{\frac{1}{2}}
\newcommand{\la}{\frac{\Lambda}{6}}
\newcommand{\G}{{\it\Gamma}^}

\newcommand{\Aa}{{\rm A}}
\newcommand{\BB}{{\rm B}}

\begin{document}
\large

 {\LARGE{\bf
 \centerline {Gravitational waves in vacuum spacetimes}
 \centerline {with cosmological constant.}
 \centerline {I. Classification and geometrical properties}
 \centerline {of non-twisting type {\it N} solutions.} }}
 \vspace{10mm}
 Ji\v r\'\i\  Bi\v c\' ak  and  Ji\v r\' \i\  Podolsk\' y
 \vspace{5mm}

 {\it
 Department of Theoretical Physics, Faculty of Mathematics and Physics,

 Charles University, V Hole\v sovi\v ck\'ach 2, 180 00 Prague 8,
 Czech Republic}
 \vspace{5mm}

 {\footnotesize
 Electronic addresses:
        bicak@mbox.troja.mff.cuni.cz,
        podolsky@mbox.troja.mff.cuni.cz}
 \vspace{10mm}

\begin{abstract}
All non-twisting Petrov-type {\it N} solutions of vacuum Einstein field
equations with cosmological constant $\Lambda$ are summarized.
They are shown to belong either to the non-expanding Kundt
class or to the expanding Robinson-Trautman class.
Invariant subclasses of each class are defined and the
corresponding metrics are given explicitly in suitable
canonical coordinates. Relations between the subclasses
and their  geometrical properties are analyzed.
In the subsequent paper these solutions are interpreted as
exact gravitational waves propagating in de~Sitter or
anti--de~Sitter spacetimes.

\vspace{3mm}
\noindent
PACS number(s): 04.30.-w, 04.20.Jb, 98.80.Hw
\end{abstract}
\newpage

\noindent
{\bf I. INTRODUCTION AND SUMMARY}
\bigskip

The purpose of this and the subsequent paper is to analyze all
non-twisting type {\it N\ } solutions of Einstein's vacuum
equations with $\Lambda$. There are several basic works on these
solutions available in literature, in particular \cite{GDP}-\cite{ORR}
(for pre-1980 works, see Ref.
\cite{KSMH}). None of them, however, discusses the physical interpretation
of the solutions. Such an interpretation, based on the study of
the deviation of geodesics,
will be presented in the following paper. In this part we
summarize, compare, classify and generalize the mathematical results
of Refs. \cite{GDP}-\cite{ORR}.

We consider  type {\it N\ } solutions in which the Debever-Penrose
null vector field {\bf k} is quadruple \cite{KSMH}.
The vector field defines a congruence of null geodesics
$x^\alpha (v)$ such that $dx^\alpha/dv=k^\alpha$,
$k_\alpha k^\alpha =0$, $k_{\alpha ; \beta}k^\beta=0$,
$v$ being an affine parameter.
In general a geodesic congruence is characterized by its
expansion $\Theta = \pul k^\alpha _{\ ;\alpha} $,
shear $\left|\sigma \right|= \sqrt{\pul k_{(\alpha;\beta)}k^{\alpha;\beta}
  - \Theta^2}$
and twist $\omega = \sqrt{\pul k_{[\alpha ; \beta]}k^{\alpha ; \beta}}$
\cite{KSMH}.
The Bianchi identities and the Kundt-Thompson theorem for
type {\it N\ } solutions (see \cite{KT}, Theorem 7.5 in \cite{KSMH})
imply $\sigma=0$ since
${C_{\alpha\beta\gamma\delta}}^{;\delta}
={R_{\alpha\beta\gamma\delta}}^{;\delta}=0 $
for solutions of $ R_{\alpha\beta}=\Lambda g_{\alpha\beta} $.
In the following we assume $\omega=0$.
Therefore, with (possibly) non-vanishing $\Lambda$ we are left
with two cases to consider:
(i) the Kundt class of non-expanding gravitational waves,
$\Theta=0$ (cf. \cite{Kundt}, Ch.27 in \cite{KSMH}),
and (ii) the Robinson-Trautman class of expanding gravitational
waves of type {\it N\ }, $\Theta\not=0$
(cf. \cite{RT}, Ch.24 in \cite{KSMH}).
Hereafter we denote the Kundt class by $KN(\Lambda)$ and the
Robinson-Trautman class by $RTN(\Lambda)$.

Section II analyzes the class $KN(\Lambda)$ in detail. We do not
present any new solution of this type but we extend the results
of Ref. \cite{ORR} by giving the explicit forms of the
transformations which leave the metrics of $KN(\Lambda)$ invariant.
These enable us to give the {\it explicit} transformations to
different canonical subclasses of $KN(\Lambda)$, which have not
been given in literature so far. We also introduce
a convenient new notation for these subclasses and demonstrate how
they are interconnected. In particular, we show that one of the
subclasses is identical to the `Lobatchewski waves' studied by
Siklos \cite{Sik}. We also formulate the proposition (proven in
Appendix) that all vacuum solutions with $\Lambda$ which are
conformal to the `Kundt waves with $\Lambda=0$' belong to one specific
subclass of $KN(\Lambda)$; we thus generalize the result
of an analogous theorem for the {\it pp}-waves \cite{Sik}.

In Section III the Robinson-Trautman solutions \cite{RT} of type {\it N\ }
with $\Lambda$ are discussed. The transformation between
two standard forms of the metric (those of Refs. \cite{KSMH} and
\cite{GDP}) is presented, and then transformations preserving the
metric form given in \cite{GDP} are generalized and  used
to demonstrate how the non-radiative part of the structural function
of these solutions can be transformed away.

\bigskip
\bigskip
\noindent
{\bf II. THE KUNDT CLASS OF SOLUTIONS {\boldmath $KN(\Lambda)$}}

\bigskip
This class has been  investigated in detail by Ozsv\'ath, Robinson
and R\'ozga \cite{ORR}. They have shown that in this case the vector {\bf k}
can be normalized such that
$\pul{\cal L}_{\bf k} g_{\alpha\beta}=\pul(k_{\alpha;\beta}
+k_{\beta;\alpha}) = Lk_\alpha k_\beta$,
where ${\cal L}_{\bf k}$ is the Lie derivative and $L$ is a scalar.
Denoting $L'={\cal L}_{\bf k} L$, we find ${\cal L}_{\bf k}L'=0$
so that $L'$ is invariant under renormalization of {\bf k}.
A suitable coordinate system $(v,\xi,\bar\xi,u)$, where
$\xi,\bar\xi$ are space-like coordinates, $v$ is a parameter along
the null geodesics and $u$ is a retarded time with $u=$ const. being
a wavefront, ${\bf k} \equiv \psi (u)\partial_v$, can be
introduced \cite{ORR} in which the $KN(\Lambda)$ metrics have the form
\begin{equation}
ds^2=2\frac{1}{p^2}d\xi d\bar\xi-2\frac{q^2}{p^2}dudv + F du^2\ ,
\label{E3.2}
\end{equation}
where
\begin{eqnarray}
&&p=1+ \la \xi\bar\xi\ , \qquad
q=(1-\la \xi\bar\xi)\alpha+\bar\beta\xi+\beta\bar\xi\ , \nonumber \\
&&F=\kappa{q^2\over p^2}v^2 - {(q^2),_u\over p^2}v - {q\over p}H\ ,
\qquad \kappa={\Lambda\over 3}\alpha^2+2\beta\bar\beta\ . \nonumber
\end{eqnarray}
Here $\alpha(u)$ and $\beta(u)$ are {\it arbitrary} real and complex functions
of $u$, respectively. These functions play the role of two arbitrary
`parameters', i.e., we can denote the Kundt class by
$KN(\Lambda)\equiv KN(\Lambda)[\alpha,\beta]$. The parameter $\kappa$  is
related to the invariant $L'$ by
\begin{equation}
L'=\kappa{p^2\over q^2}\ .
\label{E3.3}
\end{equation}
The function $H=H(\xi,\bar\xi,u)$ entering $F$ is restricted by
Einstein's equations,
$H_{,\xi\bar\xi}+(\Lambda/3p^2)\,H=0$.
There exists a general solution to this equation
\begin{equation}
H(\xi,\bar\xi,u)=(f_{,\xi}+\bar f_{,\bar\xi})-{\Lambda\over3p}
    (\bar\xi f + \xi\bar f)\ ,
\label{E3.5}
\end{equation}
where $f(\xi,u)$ is an arbitrary function of $\xi$ and $u$, analytic in
$\xi$. The space-time is conformally flat
if and only if the structural function $H$ is of the form
\begin{equation}
H=H_c={1\over p}{\Big[(1-\la \xi\bar\xi){\cal A}+
                       \bar {\cal B}\xi+{\cal B}\bar\xi\Big]}\ ,
\label{E3.6}
\end{equation}
with ${\cal A}(u)$ and ${\cal B}(u)$ being arbitrary real and complex
functions, respectively. Since $H_c$ of this form corresponds to
(\ref{E3.5}) for $f$ quadratic in $\xi$ we easily infer the
following
\bigskip

\noindent
{\it Lemma 1:}
The $KN(\Lambda)$ solutions (\ref{E3.2}), (\ref{E3.5}) with
$f=f_c= c_0(u)+c_1(u)\xi+c_2(u)\xi^2$, where $c_i(u)$ are arbitrary
complex functions of $u$, are isometric to
Minkowski (if $\Lambda=0$), de~Sitter ($\Lambda>0$) and
anti-de~Sitter spacetime ($\Lambda<0$).
\bigskip

It can be proven by straightforward but tedious
calculations that the following lemma (which is not formulated in
\cite{ORR} but is a consequence of results therein) is true:
\bigskip

\noindent
{\it Lemma 2:}
 The metric of the $KN(\Lambda)[\alpha,\beta]$ class preserves its form
 (\ref{E3.2}) under the transformations
$(v,\xi,\bar\xi,u)$ $\rightarrow (w,\eta,\bar \eta,t)$ given by
\begin{eqnarray}
&&v=a(t)\Bigg[w+{{(1-\la \eta\bar\eta)\,\gamma\>+\,\delta\,\bar\eta\,+
  \,\bar\delta\,\eta}\over{(1-\la \eta\bar\eta)
  \alpha'+\beta'\bar\eta+\bar\beta'\eta}}\Bigg]+\Delta(t)\ , \nonumber \\
&&\xi={{\bar B(t)+A(t)\eta}\over{\bar A(t)-\la B(t)\eta}}
\ ,\qquad u=u(t)\ ,\label{E3.8}
\end{eqnarray}
where $A(t), B(t)$ are arbitrary complex and $a(t), u(t)$ are real
functions of $t$, respectively.
\bigskip

In the new coordinates $(w,\eta,\bar\eta,t)$
the resulting metric $KN(\Lambda)[\alpha',\beta']$ has
\begin{eqnarray}
\alpha'&=&{{\sqrt{a\dot u}}\over\Phi}\left[{(A\bar A-\la B \bar B)\alpha
                      +\bar A B\beta+A\bar B \bar \beta}\right]\ , \nonumber \\
\beta'&=&{{\sqrt{a\dot u}}\over\Phi}\left[{-{\Lambda\over 3}\bar A \bar B
                 \alpha +\bar A^2 \beta -\la \bar B^2 \bar
                \beta}\right]\ ,   \label{E3.10}
\end{eqnarray}
with $\Phi=A \bar A+\la B \bar B$ and the dot denotes $d/dt$.
The remaining real functions $\gamma(t) , \Delta(t)$ and a complex function
$\delta(t)$ in (\ref{E3.8}) must satisfy the equations
\begin{eqnarray}
\alpha'\delta-\beta'\gamma&=&{1\over \Phi}(\bar A \dot{\bar B}
              - \dot{\bar A} \bar B)\equiv C\ , \nonumber \\
\bar \beta' \delta-\beta' \bar \delta&=&{1\over \Phi}
              (\dot A \bar A-A \dot{\bar A}+{\la}\dot B \bar B
              -{\la}B \dot{\bar B})\equiv D \ , \label{E3.11} \\
{{\kappa'}\over a}\Delta&=&\pul({\dot a\over a}-{\ddot u \over \dot u})
                 -({\Lambda\over 3}\alpha' \gamma+\bar \beta' \delta
           +\beta' \bar \delta)\ . \nonumber
\end{eqnarray}
The structural function $H'(\eta ,\bar \eta, t)$ then takes the form
\begin{equation}
p'H'=(1-\la \eta \bar \eta)E+\bar F \eta+F \bar \eta
     +{\dot u\over \Phi}{\sqrt{\dot u\over a}}(A-\la \bar B \bar \eta)
       (\bar A -\la B\eta)pH\ ,\label{E3.12}
\end{equation}
where
\begin{eqnarray}
E(t)&\equiv& 2\dot \gamma+{2\over a}(\alpha' \Delta)^\cdot-\alpha' \kappa'{\Delta^2\over a^2}
      -2{\Delta\over a}\left[{{1\over 2}\alpha'({\dot a\over a}+{\ddot u\over \dot u})
       +C\bar \beta' + \bar C \beta'}\right] \nonumber\\
    &\ & -\alpha'(2\delta\bar\delta
    -{\Lambda\over 3}\gamma^2)+2\gamma(\beta'\bar \delta+
          \bar \beta'\delta)\ , \nonumber\\
F(t)&\equiv& 2\dot \delta+{2\over a}(\beta' \Delta)^\cdot-\beta' \kappa'{\Delta^2\over a^2}
      -2{\Delta\over a}\left[{{1\over 2}\beta'({\dot a\over a}+{\ddot u\over \dot u})
       -{\Lambda\over 3}C\alpha' - D \beta'}\right] \nonumber\\
   &\ & -{\Lambda\over 3}\gamma(\beta' \gamma-2\alpha' \delta)
     +2\bar \beta' \delta^2\ .\nonumber
\end{eqnarray}

The  transformation (\ref{E3.8}) is also important in connection
with Lemma 1. Comparing Eq. (\ref{E3.6}) with Eq. (\ref{E3.12}),
we see that we can `generate' \ $H=H_c$ from $H=0$ by a coordinate
transformation. Therefore, the conformally flat part, $H_c$, of  $H$
generated by  $f_c$ cannot represent
a radiative field. Only if the function $f$ is at least cubic in
$\xi$, the resulting spacetime is of type {\it N\ } and can be
interpreted as radiative (see the subsequent paper).

Another important application of the coordinate freedom
(\ref{E3.8})  was suggested in \cite{ORR}: it is possible to use
the transformation to get `canonical'  subclasses of
$KN(\Lambda)[\alpha, \beta]$ corresponding to special values of
parameters $\alpha$ and $\beta$. Without loss of generality we can assume
$\alpha\geq 0$. In addition, the transformation (\ref{E3.8}) of
a special form
\begin{equation}
\xi=\eta\ ,\quad
  v=a(t) w\ ,\quad
  u=\int a(t)\,dt\ ,\label{E3.13}
\end{equation}
for $a(t)\not =0$ results in  scaling
$\alpha(u)=\alpha'(t)/|a(t)|$,
$\beta(u) =\beta'(t)/|a(t)|$
so that {\it we can always assume either $\alpha=1$ or $\alpha=0$}.
Since the parameter $L'$ (\ref{E3.3}) is invariant,
the sign of $\kappa$ is also an invariant. We can
thus base the invariant canonical classification of all $KN(\Lambda)$
solutions on the sign of $\kappa$ and the sign of $\Lambda$.
There are nine  possible cases ($\kappa$ and $\Lambda$ can both be positive,
zero or negative). However, subclasses $\kappa<0 , \Lambda>0$  and
$\kappa=0 , \Lambda>0$  and  $\kappa<0 , \Lambda=0$
are forbidden since they violate the relation
$\kappa={\Lambda\over 3}\alpha^2+2\beta\bar\beta$.
The remaining possibilities are six subclasses which we shall now
discuss.
\medskip

\noindent
{\bf A. Subclass $\kappa=0 , \Lambda=0$.}
 The equation $\kappa=0$ implies  $\beta=0$ so that $\alpha=1$.
 A canonical representative of this subclass can be denoted as
 $PP\equiv KN(\Lambda=0)[\alpha=1,\beta=0]$.
 We are now using the notation  $PP$  since the corresponding metric
\begin{equation}
PP:\qquad ds^2=2d\xi d\bar\xi - 2dudv - (g+\bar g)du^2\ ,  \label{E3.14}
\end{equation}
 with arbitrary $g(\xi,u)=f_{,\xi}$ analytic in $\xi$, describes
 well-known {\it pp} waves investigated by many authors
 (for details and references see section 21.5 in \cite{KSMH}).
\medskip

\noindent
{\bf B. Subclass $\kappa>0 , \Lambda=0$.}
  Since $0<\kappa=2\beta\bar\beta$, we have $\beta\not =0$.
  Then Eqs. (\ref{E3.8}) with
$A=\sqrt\beta$, $B=(-\pul{\alpha\over \beta}+i{J\over\beta})\,
A$,  $a={1\over\beta\bar\beta}$,
$\gamma=\pul\Big({\alpha\over{\sqrt{\beta\bar\beta}}}\Big)^\cdot
+{2J\over\alpha}\Big({J\over{\sqrt{\beta\bar\beta}}}\Big)^\cdot$,
$\delta=\dot A/A$,  $\Delta=\pul\dot a$, $u=t$, where $J$ is a
real function satisfying $\Big({J\over{\sqrt{\beta\bar\beta}}}\Big)^\cdot
={i\over4}{\alpha\over{\sqrt{\beta\bar\beta}}}{\bar\beta\over\beta}
\Big({\beta\over\bar\beta}\Big)^\cdot$,
 transform any solution of this subclass to
 $KN\equiv KN(\Lambda=0)[\alpha=0,\beta=1]$.
 The metric
\begin{equation}
KN:\qquad ds^2=2d\xi d\bar\xi +2(\xi+\bar\xi)^2\left[-dv+
   \left(v^2-\frac{g+\bar g}{\xi+\bar\xi}\right)du\right]du\ , \label{E3.15}
\end{equation}
 with arbitrary $g(\xi,u)=\pul f_{,\xi}$, describes  solutions discovered
 by Kundt \cite{Kundt}.  This is a {\it special}
 `Kundt solution' in the `Kundt class' of non-expanding waves.
 For details see \cite{KSMH}, Chapter 27;
 the transformation between $\tilde v$ used there and $v$ above
 is $v={{\tilde v}/{(\xi+\bar\xi)^2}}$.
\medskip

\noindent
{\bf C. Subclass $\kappa>0 , \Lambda>0$.}
If $\alpha=1$ we can use (\ref{E3.8}) with $A=\exp {(-{i\over2}\chi)}$,
$\ B=cA$, $\ a=(\la+\beta\bar\beta)^{-1}$, $\ \gamma=-\Big[\dot W+
{1\over2}\Big(\dot V\exp{(-i\chi)}+\dot{\bar V}\exp{(i\chi)}\Big)\Big]
/(1+\la c\bar c)$,
$\ \delta=\dot A/A+\la\dot c\bar c/(1+\la c\bar c)$,
$\ \Delta={1\over4}\dot a-{\Lambda\over12}{{(c\bar c)}^\cdot(1+\la c\bar
c})^{-1}a$, $\ u=t$, where $c\equiv V+W\bar A/A$,
$\ V\equiv {6\over\Lambda}\bar \beta$,
$\ W\equiv -{6\over\Lambda}\sqrt{\la+\beta\bar\beta}$,
$\ \chi$ is a real function satisfying $\dot\chi={i\over 2}
\Big(\dot V\exp{(-i\chi)}-\dot{\bar V}\exp{(i\chi)}\Big)/W$ to
obtain $\alpha'=0$, $\beta'=1$.
If $\alpha=0$ the scaling (\ref{E3.13}) makes
$|\beta|=1$ and then we put
$\xi={6\over\Lambda}\beta/\eta$,
$v=w+\pul(\dot\beta\bar\beta\eta+\beta\dot{\bar\beta}\bar\eta)/
(\eta+\bar\eta)$,
$u=t$.  In both cases the representative of this subclass is
$KN(\Lambda)I\equiv KN(\Lambda)[\alpha=0,\beta=1]$.
The metric reads

 $\qquad KN(\Lambda)I$:
\begin{equation}
ds^2=2{{d\xi d\bar\xi}\over{{(1+\la\xi\bar\xi)}^2}}-
   2{\left({{\xi+\bar\xi}\over{1+\la\xi\bar\xi}}\right)}^2 dudv+
   \left[ { 2{\left({{\xi+\bar\xi}\over{1+\la\xi\bar\xi}}\right)}^2 v^2 -
            {{\xi+\bar\xi}\over{1+\la\xi\bar\xi}}H} \right] du^2 \ .
 \label{E3.16}
\end{equation}
 $KN(\Lambda)I$ indicates that this solution  is a generalization
 of the $KN$ waves  to the case $\Lambda\not=0$, {\it `I'} means
 `of the first kind'.  The $KN(\Lambda)I$ solutions were first discovered
 by Garc\' \i a  D\' \i az and Pleba\' nski \cite{GDP},\cite{Gar},
 transformation between their coordinates and those
 used here is  $\xi=\sqrt{6\over\Lambda}\tanh{\sqrt{\la}\tilde\xi}$ ,
 $v={r/{\sinh{\sqrt{\la}(\tilde\xi+\bar{\tilde\xi})}}}$ , $u=-\sqrt{\la}t$.
 However,  in \cite{GDP},\cite{Gar} not all $KN(\Lambda\not=0)$
 solutions were found since (invariantly different) subclasses E and F
 mentioned below were omitted.
\medskip

\noindent
{\bf D. Subclass $\kappa>0 , \Lambda<0$.}
  The same transformation as in the previous case leads to
 the metric (\ref{E3.16}) which
  has thus the same form  for  $\Lambda<0$.
 The transformation to the coordinates used in \cite{GDP} is
 $\xi=\sqrt{-{6\over\Lambda}}\tan{\sqrt{-\la}\tilde\xi}$ ,
 $v={r/{\sin{\sqrt{-\la}(\tilde\xi+\bar{\tilde\xi})}}}$ , $u=-\sqrt{-\la}t$.
\medskip

\noindent
{\bf E. Subclass $\kappa<0 , \Lambda<0$.}
 Now $\kappa<0$ implies  $\alpha=1$.
 Using (\ref{E3.8}) with $A=\exp {(i\phi)}$,  $B=cA$,
$a=(1+{6\over\Lambda}\beta\bar\beta)^{-1}$, $\gamma=0$,
$\delta=\bar A^2\dot{\bar c}/(1+\la c\bar c)$,
$\Delta={3\over2\Lambda}\dot a$, $u=t$,
where $c\equiv {1\over\beta}\Big(\sqrt{1+{6\over\Lambda}
\beta\bar \beta} -1\Big)$, $\phi$ is a real function satisfying
$\dot\phi=i\frac{\Lambda}{12}
(\dot c\bar c-c\dot{\bar c})/( 1+\la c\bar c)$,
we find $\beta'=0$ and the canonical representative,
denoted by $KN(\Lambda^-)II\equiv KN(\Lambda<0)[\alpha=1,\beta=0]$,
is

$\qquad KN(\Lambda^-)II$:
\begin{equation}
ds^2=2{{d\xi d\bar\xi}\over{{(1+\la\xi\bar\xi)}^2}}-
   2{\left({{1-\la\xi\bar\xi}\over{1+\la\xi\bar\xi}}\right)}^2 dudv+
   \left[ { {\Lambda\over3}{\left({{1-\la\xi\bar\xi}\over{1+\la\xi\bar\xi}}\right)}^2 v^2 -
            {{1-\la\xi\bar\xi}\over{1+\la\xi\bar\xi}}H} \right]
du^2\ . \label{E3.17}
\end{equation}
 Here $KN(\Lambda^-)II$ means generalized Kundt waves
 `of the second kind' with $\Lambda<0$. This class
 was first discovered by Ozsv\' ath {\it et.al.}
\cite{ORR}. Observe  that with $\kappa>0$,
$\Lambda>0$,  $KN(\Lambda^+)II\equiv KN(\Lambda^+)I$,  since  transformation
 (\ref{E3.8})  where $A={1\over\sqrt 2}$, $B=\sqrt {3\over\Lambda}$, $a=1$,
 $\gamma=\delta=\Delta=0$, $u={\Lambda\over 6} t$ identifies
 $KN(\Lambda^+)[\alpha=0,\beta=1]$ with $KN(\Lambda^+)[\alpha'=1,\beta'=0]$.
\medskip

\noindent
{\bf F. Subclass $\kappa=0 , \Lambda<0$.}
 The relation $\kappa=0$ implies
 $\alpha=1$. Thus, $|\beta|=\sqrt{-\la}$
 and the representative can be denoted by $KN(\Lambda^-)III\equiv
   KN(\Lambda<0)[\alpha=1,\beta=\sqrt{-\la}e^{i\omega(u)}]$,
$\omega(u)$ being an arbitrary real function of $u$. The metric takes the form
of  (\ref{E3.2}) with $q=\Big(1+\sqrt{-\la}\xi e^{-i\omega(u)}\Big)
\Big(1+\sqrt{-\la}\bar\xi e^{i\omega(u)}\Big)$. This class
 was also first discovered in \cite{ORR}.
  One can distinguish {\it two subsubclasses} of
 $KN(\Lambda^-)III$ according to the value of $L$:
1.  $\ L\not=0$ ({\bf k} is not a Killing vector).
2.  $\ L=0$ ({\bf k} is a Killing vector).
In case 2. $\beta=$const. and
the transformation $\xi=\sqrt{\beta/\bar\beta}\,\eta$ leads to
$KN(\Lambda^-)III_K\equiv KN(\Lambda<0)[\alpha=1,\beta=\sqrt{-\la}]$,
where the suffix {\it `K'} stands for `Killing':
\begin{eqnarray}
KN(\Lambda^-)III_K:\quad &&ds^2=2{{d\xi d\bar\xi}\over{{(1+\la\xi\bar\xi)}^2}}-
  2{\left({{(1+\sqrt{-\la}\xi)(1+\sqrt{-\la}\bar\xi)}
    \over{1+\la\xi\bar\xi}}\right)}^2 dudv \nonumber\\
 &&\hskip30mm-{{(1+\sqrt{-\la}\xi)(1+\sqrt{-\la}\bar\xi)}\over{1+\la\xi\bar\xi}}H du^2
\ .\label{E3.18a}
\end{eqnarray}
This subsubclass can be shown to be identical with the `Lobatchevski waves'
studied by Siklos \cite{Sik}: indeed, the transformation
$\xi=-\sqrt{-{6\over\Lambda}}(x+{1\over2}+iy)/(x-{1\over2}+iy)$,
$v={12\over\Lambda}\,r$,
brings the metric (\ref{E3.18a}) into the form
\begin{equation}
ds^2=-{3\over{\Lambda}}\cdot{1\over{x^2}}\left(dx^2+dy^2+2dudr+\tilde H du^2\right)
\ ,\label{E3.18c}
\end{equation}
where $\tilde H\equiv-\la xH$. This is the Siklos metric
\cite{Sik}. Recently we analyzed in detail the behaviour of
test particles in these solutions and interpreted them as waves
in the anti-de~Sitter spacetime \cite{JiPo}. Impulsive waves of
this type were investigated in \cite{PoGriff}.
\medskip

\noindent
{\bf G. Relations of the subclasses.}
Using the above results we can summarize the invariant canonical
classification of the $KN(\Lambda)$ class of solutions in the following diagram:
\[ KN(\Lambda)\ \left\{\begin{array}{ll}
            \ \Lambda=0\      \left\{\begin{array}{l}
                       \kappa=0: PP  \\
                       \kappa>0: KN  \\
                       \end{array}\right. \\
                       &\\
            \ \Lambda\not=0\  \left\{\begin{array}{l}
                       \kappa>0: KN(\Lambda)I   \\
                       \kappa<0: KN(\Lambda^-)II  \\
                       \kappa=0: KN(\Lambda^-)III \rightarrow
                                 KN(\Lambda^-)III_K\ . \end{array}\right. \\
             \end{array}\right. \]
There is an asymmetry with respect to the sign of $\Lambda$:
there are {\it three} distinct classes of non-expanding waves
for $\Lambda<0$  whereas there is only {\it one} such  class for $\Lambda>0$.
The reason is in the condition $\kappa={\Lambda\over3}\alpha^2+2\beta\bar\beta$
which for $\Lambda>0$ excludes the cases $\kappa<0$ and $\kappa=0$.
Intuitively, fewer non-expanding waves `fit' into the
de~Sitter universe which admits closed spacelike sections than
into the anti-de~Sitter space.

There exist natural relations between the $\Lambda=0$
and $\Lambda\not=0$ subclasses. The metrics (\ref{E3.16}),
(\ref{E3.17}) and (\ref{E3.18a}) do not diverge as
$\Lambda\rightarrow0$, we can set $\Lambda=0$ and thus find
\begin{equation}
KN(\Lambda=0)I=KN\ , \quad
KN(\Lambda^-=0)II=PP\ , \quad
KN(\Lambda^-=0)III=PP\ .   \label{E3.19}
\end{equation}
Thus, it is natural to consider the $KN(\Lambda)I$ class as a
generalization of the Kundt solution $KN$, and the
classes $KN(\Lambda^-)II$  and $KN(\Lambda^-)III$ as generalizations of
$PP$ waves. There exists {\it no} generalization of  $PP$ waves to the case
of $\Lambda>0$.

From metrics (\ref{E3.14}) and (\ref{E3.18c}), with
$\xi={(1/\sqrt2)}(x+iy)$, and from (\ref{E3.15}) and (\ref{E3.16}) we find
$ds^2_{KN(\Lambda^-)III_K}=-{6\over{\Lambda}}(\xi+\bar\xi)^{-2}ds^2_{PP}$,
and $ds^2_{KN(\Lambda)I}=(1+\la\xi\bar\xi)^{-2}ds^2_{KN}$.
Therefore, the class
$KN(\Lambda^-)III_K$  is conformal to the $PP$-class, and the
class $KN(\Lambda)I$ is conformal to the $KN$-class.
In fact, the solutions  $KN(\Lambda^-)III_K$ and $KN(\Lambda)I$
are the only non-trivial space-times conformal to $PP$ and $KN$,
respectively. The theorem proven by Siklos \cite{Sik} states that
the only vacuum solutions (other than $PP$ solutions themselves)
which are properly (with non-constant factor)
conformal to non-flat $PP$ metrics are $KN(\Lambda^-)III_K$ metrics.
However, we see from (\ref{E3.19}) that $PP$
metrics are a special case of metrics $KN(\Lambda^-)III_K$ for
$\Lambda=0$, and  Siklos' theorem may just be formulated as
follows:
\smallskip

{\it Proposition 1:}
The only vacuum solutions conformal to non-flat $PP$ metrics
are $KN(\Lambda^-)III_K$ metrics.
\bigskip

\noindent
In addition, the following analogous proposition can be proven
for the $KN$ solutions:
\bigskip

{\it Proposition 2:}
The only vacuum solutions conformal to $KN$ metrics are $KN(\Lambda)I$
metrics.
\bigskip

\noindent
The proof is contained in Appendix.

The conformal, homothetic, and isometric symmetries of the $KN(\Lambda)$
solutions have been systematically investigated by Salazar, Garc\'\i a
and Pleba\'nski \cite{SGDP} and by Siklos \cite{Sik}.
It is only in the subclasses $PP$ and $KN(\Lambda^-)III_K$  that the vector
${\bf k}=\partial_v$  is a Killing vector;
$k_{\alpha;\beta}=0$ only in the $PP$ subclass.
Let us finally summarize the classification of all $KN(\Lambda)$ solutions
and compare our notation with notations used in literature:

\bigskip
\begin{tabular}{|l|l|}
\hline
 {\it Notation in this paper} &
 {\it Notations in literature} \\
\hline
  $KN(\Lambda)$ & $R(\Lambda,\alpha,\beta)$\ \  \cite{ORR}\\
\hline
  $PP$         & $pp$\ ,\ $R$\ \  \cite{KSMH}, \cite{ORR} for Refs. \\
\hline
  $KN$         & $K$\ \  \cite{Kundt}, \cite{KSMH}\\
\hline
  $KN(\Lambda)I$& $K(\Lambda)$\ \  \cite{GDP}, \cite{ORR}\\
\hline
  $KN(\Lambda^-)II$ & $R(\Lambda)$\ \  \cite{ORR}\\
\hline
  $KN(\Lambda^-)III$& $(IV)_1$\ \  \cite{ORR}\\
\hline
  $KN(\Lambda^-)III_K$&$(IV)_0$\ \  \cite{Sik}, \cite{ORR}\\
\hline
\end{tabular}
\bigskip

\noindent
Impulsive waves in the $KN(\Lambda)$ spacetimes were
recently studied in \cite{JP}.

\bigskip
\bigskip

\noindent
{\bf III. THE ROBINSON-TRAUTMAN CLASS OF SOLUTIONS
         {\boldmath $RTN(\Lambda,\epsilon)$}}

\bigskip
The Robinson-Trautman solutions \cite{RT} satisfying the vacuum
equations with $\Lambda$ can be written as (see \cite{KSMH})
\begin{equation}
ds^2=2{r^2\over P^2}d\zeta d\bar\zeta-2dudr-\Big[\Delta\ln P-2r(\ln P)_{,u}-
       {2m\over r}-{\Lambda\over3}r^2\Big]du^2\ , \label{E4.25}
\end{equation}
where $\zeta$ is a complex spatial coordinate, $r$ is an affine parameter
along the rays generated by the vector field
{\bf k}, $u$ is a retarded time, $m$ is a function of $u$ which in some
cases can be interpreted as mass,
and $\Delta\equiv2P^2\partial^2/\partial\zeta\partial\bar\zeta$.
The function $P\equiv P(\zeta,\bar\zeta,u)$ satisfies the equation
$\Delta\Delta(\ln P)+12m(\ln P)_{,u}-4m_{,u}=0$.
Here we restrict attention to the solutions of type {\it N\ }
and denote these as $RTN(\Lambda)$.
In this case  $m=0$ and $\Delta\ln P= K(u)$.
By a transformation
$u=g(\tilde u)$, $r={\tilde r/\dot g}$,  where $\dot g={dg/ d\tilde u}$,
we can set the Gaussian curvature $K(u)$ of the 2-surfaces
$2P^{-2}{d\zeta d\bar\zeta}$ to be $K=2\epsilon$, where
$\epsilon=+1,0,-1$ (since $\tilde P=\dot g P$
and $\tilde K={\dot g}^2 K$, the sign of $K$ is invariant).
Thus, the different subclasses  can be denoted as
$RTN(\Lambda,\epsilon)$. The corresponding
metrics can be written as
\begin{equation}
ds^2=2{r^2\over P^2}d\zeta d\bar\zeta-2dudr-2\Big[\epsilon-r(\ln P)_{,u}
       -{\Lambda\over6}r^2\Big]du^2\ .\label{E4.29}
\end{equation}
Since $\epsilon=+1,0,-1$ and $\Lambda>0$, $\Lambda=0$,
$\Lambda<0$, there are 9 invariant subclasses.

Another coordinates for the $RTN(\Lambda,  \epsilon)$
class, suitable for physical interpretation, has been given
by Garc\'\i a D\'\i az and Pleba\'nski \cite{GDP}.
Their metric is expressed in terms of a function $f(\xi,u)$ which is an
arbitrary function of $u$, analytic in spatial coordinate $\xi$
\begin{equation}
ds^2=2v^2d\xi d\bar\xi+2v\bar {\Aa} d\xi du+2v\Aa d\bar\xi du
 + 2 \psi dudv+2(\Aa\bar {\Aa}+\psi \BB)du^2\ , \label{E4.30}
\end{equation}
where
$$\Aa=\epsilon\xi-v f\ ,\qquad
\BB=-\epsilon+{v\over2}(f_{,\xi}+\bar f_{,\bar\xi})+\la
     v^2\psi\ ,\qquad
\psi=1+\epsilon\xi\bar\xi\ .$$
It can be shown that the transformation
relating (\ref{E4.29}) with (\ref{E4.30}) has the form
\begin{eqnarray}
\xi&=&F(\zeta,u)=\int f\Big(\xi(\zeta,u),u\Big)\,du\ ,
          \nonumber\\
v  &=&{r\over 1+\epsilon F\bar F}\ , \label{E4.31}\\
u &\rightarrow& -u\ ,\qquad\quad
P=(1+\epsilon F\bar F)(F_{,\zeta} \bar F_{,\bar\zeta})^{-\pul}
\ . \nonumber
\end{eqnarray}
If $f$ does not depend on $\xi$, we put $\xi=F(\zeta,u)=\zeta+\int f(u)\,du$.

The non-vanishing Weyl tensor components are proportional to $f_{,\xi\xi\xi}$
so that the solutions are conformally flat if $f$ is quadratic in $\xi$.
Thus, we can formulate
\bigskip

\noindent
{\it Lemma 3:}
The $RTN(\Lambda, \epsilon)$ solutions (\ref{E4.30}) with
$f=f_c= c_0(u)+c_1(u)\xi+c_2(u)\xi^2$, where $c_i(u)$ are arbitrary
complex functions of $u$, are isometric to
Minkowski (if $\Lambda=0$), de~Sitter ($\Lambda>0$)  and
anti-de~Sitter spacetime ($\Lambda<0$).
\bigskip

Transformations preserving the form of (\ref{E4.30}) were
studied in \cite{SGDP}. However, more general transformations
(Eq. (2.4) in \cite{SGDP} follows from Eqs. (\ref{E4.33a}),
(\ref{E4.36a}) if we put $\Delta=0$ and
$\alpha\bar\alpha+\epsilon\beta\bar\beta=1$) are given in the following
\bigskip

\noindent
{\it Lemma 4:}
 The coordinate transformations $(v,\xi,\bar\xi,u)\rightarrow
(w,\eta,\bar\eta,t)$, which maintain invariant the form of the
$RTN(\Lambda,  \epsilon)$ metric (\ref{E4.30}), are
\begin{eqnarray}
\xi&=&\frac{\alpha\eta+\beta}{\gamma\eta+\delta}\ , \nonumber \\
v&=&\frac{(\gamma\eta+\delta)(\bar\gamma\bar\eta+\bar\delta)}{\sqrt
   {(\alpha\delta-\beta\gamma)(\bar\alpha\bar\delta-\bar\beta\bar\gamma)}}\,w
   \ ,                                             \label{E4.33a} \\
u&=&\int \frac{\sqrt{(\alpha\delta-\beta\gamma)
                 (\bar\alpha\bar\delta-\bar\beta\bar\gamma)}}
              {\delta\bar\delta+\epsilon\beta\bar\beta}\,dt
   \ , \nonumber
\end{eqnarray}
where $\alpha(t),\beta(t),\gamma(t),\delta(t)$ are arbitrary complex
functions of $t$ which satisfy the conditions
\begin{equation}
\bar\gamma\delta=-\epsilon\bar\alpha\beta\ , \qquad
\gamma\bar\gamma+\epsilon\alpha\bar\alpha=\epsilon(\delta\bar\delta+
 \epsilon\beta\bar\beta) \ . \label{E4.33b}
\end{equation}
In the coordinates $(w,\eta,\bar\eta,t)$, new structural
function $f'(\eta,t)$ is related to $f$ as follows:
\begin{eqnarray}
f'&=&\frac{\bar\alpha\bar\delta-\bar\beta\bar\gamma}
        {\delta\bar\delta+\epsilon\beta\bar\beta}
   \frac{(\gamma\eta+\delta)^2}
         {\sqrt{(\alpha\delta-\beta\gamma)
                (\bar\alpha\bar\delta-\bar\beta\bar\gamma)}}\,f
      \nonumber\\
&&-\frac{1}{\alpha\delta-\beta\gamma}
\left[{(\dot\beta\delta-\beta\dot\delta)+
     (\dot\alpha\delta-\alpha\dot\delta+\dot\beta\gamma-\beta\dot\gamma)\eta+
     (\dot\alpha\gamma-\alpha\dot\gamma)\eta^2}\right]\ .\label{E4.35}
\end{eqnarray}
From this relation and  Lemma 3  we see that the conformally flat part
$f_c$ is indeed unimportant since it can be
generated  from $f=0$ by the coordinate transformation
(\ref{E4.33a}).

If $\epsilon\not=0$,  Eq.  (\ref{E4.33b}) implies that
$|\alpha|=|\delta|$, $|\beta|=|\gamma|$ so that
$\sqrt{(\alpha\delta-\beta\gamma)(\bar\alpha\bar\delta-\bar\beta\bar\gamma)}
 = \delta\bar\delta+\epsilon\beta\bar\beta$. Equations
(\ref{E4.33a})-(\ref{E4.35}) then simplify to
\begin{equation}
\xi={\alpha\eta+\beta\over \gamma\eta+\delta}\ ,\quad
v= {{(\gamma\eta+\delta)(\bar\gamma\bar\eta+\bar\delta)}\over
   {\alpha\bar\alpha+\epsilon\beta\bar\beta}}\,w\ ,\quad
u= t\ ,\label{E4.36a}
\end{equation}
where $\alpha(t)$ and $\beta(t)$ are arbitrary complex functions of $t$
and $\gamma(t)=\bar\beta(t) e^{i\Gamma(t)}$,
$\delta(t)=\bar\alpha(t)e^{i\Delta(t)}$,
with $\Gamma(t)$, $\Delta(t)$ being arbitrary real functions of $t$ satisfying
$\Delta-\Gamma=(1+\epsilon){\pi\over2}$.
The structural function is given by
$f'= \Big\{ (\gamma\eta+\delta)^2\,f
-[(\dot\beta\delta-\beta\dot\delta)+
     (\dot\alpha\delta-\alpha\dot\delta+\dot\beta\gamma-\beta\dot\gamma)\eta+
     (\dot\alpha\gamma-\alpha\dot\gamma)\eta^2] \Big\}
e^{-i\Delta}/(\alpha\bar\alpha+\epsilon\beta\bar\beta)$.

For $\epsilon=0$, equations (\ref{E4.33b}) imply $\gamma=0$ so that
the transformations (\ref{E4.33a}) yield
\begin{equation}
\xi=A(t)\eta+B(t) \ ,\quad
v=\frac{w}{\sqrt{A(t)\bar A(t)}}\ ,\quad
u=\int \sqrt{A(t)\bar A(t)}\, dt\ ,\label{E4.38}
\end{equation}
where $A(t)$ and $B(t)$ are arbitrary complex functions of $t$.
The relation (\ref{E4.35}) between the structural functions is
now
$f'=\sqrt{\bar A/A}\,f-(\dot B/A+\dot A/A\, \eta)$,
so that, in contrast to the case $\epsilon\not=0$, the term
quadratic in $\eta$ vanishes. Hence, Eq. (\ref{E4.38})
does not enable us to transform away the complete conformally
flat part $f_c$. (We tried but without success to generalize
(\ref{E4.38}) so that quadratic terms could be removed.)

Symmetries of the $RTN(\Lambda,  \epsilon)$ solutions have been investigated
in \cite{CF} (see also \cite{KSMH}, Table 33.2) and, more systematically,
in \cite{SGDP}. The solutions which are not conformally flat allow the existence
of at most two Killing vectors.

\bigskip
\noindent
{\bf ACKNOWLEDGMENTS}
\bigskip

\noindent
We thank Jerry Griffiths for reading the manuscript and useful
suggestions. We also acknowledge the support of grant GACR-202/99/0261
from the Czech Republic.
\bigskip

\noindent
\appendix{{\bf APPENDIX: PROOF OF PROPOSITION 2}}

\noindent
For the conformally related metrics $g_{\alpha\beta}$ and
$\hat g_{\alpha\beta}$,
$\hat g_{\alpha\beta}=\Omega^{-2}g_{\alpha\beta}$,
the trace-free Ricci tensors are related by \cite{KSMH}
\begin{eqnarray}
S_{\alpha\beta}&=&R_{\alpha\beta}-{1\over4}Rg_{\alpha\beta}
   \ ,\nonumber\\
\hat S_{\alpha\beta}&=&\hat R_{\alpha\beta}-{1\over4}\hat R\hat g_{\alpha\beta}
  =  S_{\alpha\beta}+{2\over\Omega}(\Omega_{;\alpha\beta}-
           {1\over4}g_{\alpha\beta}\Box\Omega)
\ ,\label{C3.21b}
\end{eqnarray}
where $\Box\Omega=g^{\mu\nu}\Omega_{;\mu\nu}$, the covariant derivative
is taken with respect to $g_{\alpha\beta}$, and the scalar curvatures
is
\begin{equation}
\hat R=\Omega^2 R+6\Omega\Box\Omega-12g^{\alpha\beta}
  \Omega_{;\alpha}\Omega_{;\beta}\ .\label{C3.21c}
\end{equation}
Since $\hat R_{\alpha\beta}=\Lambda \hat g_{\alpha\beta}$,
$\hat R=4\Lambda$, we have $\hat S_{\alpha\beta}=0$.
In coordinates $(v,\xi,\bar\xi,u)$,
\begin{equation}
  g_{12}=1\ ,\quad
  g_{03}=-(\xi+\bar\xi)^2  \ ,\quad
  g_{33}=2(\xi+\bar\xi)^2\Big(v^2-\pul{H\over {\xi+\bar\xi}}\Big)
  \ ,\label{C3.22}
\end{equation}
$R_{33}=(\xi+\bar\xi)\,H_{,\xi\bar\xi}$,
and the other Ricci tensor components vanish. Therefore, $R=0$
and (\ref{C3.21b}) can be written as
\begin{equation}
\Omega_{,\alpha\beta}-\G\gamma_{\alpha\beta}\Omega_{,\gamma}-
  {1\over4}g_{\alpha\beta}\Box\Omega+\pul\Omega R_{\alpha\beta}=0
  \ ,\label{C3.23}
\end{equation}
which gives
\begin{eqnarray}
&&\left({\Omega_{,v}\over {\xi+\bar\xi}}\right)_{,\xi}=0=
  \left({\Omega_{,v}\over {\xi+\bar\xi}}\right)_{,\bar\xi}
               \ ,\label{C3.24ab}\\
&&\Omega_{,vv}=0
               \ ,\label{C3.24c} \\
&&\Omega_{,\xi\xi}=0=
  \Omega_{,\bar\xi\bar\xi}
               \ ,\label{C3.24de}\\
&&\Omega_{,vu}+2v\Omega_{,v}-(\xi+\bar\xi)(\Omega_{,\xi}+\Omega_{,\bar\xi})
  +{1\over4}(\xi+\bar\xi)^2\Box\Omega=0
               \ ,\label{C3.24f} \\
&&\Omega_{,\xi\bar\xi}={1\over4}\Box\Omega
               \ ,\label{C3.24g} \\
&&\Omega_{,\xi u}-{\Omega_{,u}\over{\xi+\bar\xi}}-
    \pul\left({H\over {\xi+\bar\xi}}\right)_{,\xi}\Omega_{,v}=0=
  \Omega_{,\bar\xi u}-{\Omega_{,u}\over{\xi+\bar\xi}}-
    \pul\left({H\over {\xi+\bar\xi}}\right)_{,\bar\xi}\Omega_{,v}
               \ ,\label{C3.24hi}\\
&&\Omega_{,uu}-2v\Omega_{,u}-\left[2v\left(2v^2-{H\over {\xi+\bar\xi}}\right)
  +\pul\left({H\over {\xi+\bar\xi}}\right)_{,u}\right]\Omega_{,v}
     \nonumber\\
&&\quad\qquad-\pul[(\xi+\bar\xi)H]_{,\bar\xi}\Omega_{,\xi} -\pul[(\xi+\bar\xi)H]_{,\xi}\Omega_{,\bar\xi}
  +2v^2(\xi+\bar\xi)(\Omega_{,\xi}+\Omega_{,\bar\xi})
     \nonumber\\
&&\quad\qquad-\pul (\xi+\bar\xi)^2\left(v^2-\pul{H\over{\xi+\bar\xi}}\right)
  \Box\Omega +\pul(\xi+\bar\xi)H_{,\xi\bar\xi}\Omega=0
   \ .\label{C3.24j}
\end{eqnarray}
By using (\ref{C3.24g}), Eq. (\ref{C3.21c}) takes the form
\begin{equation}
\Omega\Omega_{,\xi\bar\xi}-\Omega_{,\xi}\Omega_{,\bar\xi}+{\Omega_{,v}
  \over{(\xi+\bar\xi)^2}}   \left[\Omega_{,u}+\left(v^2-
 \pul{H\over {\xi+\bar\xi}}\right) \Omega_{,v}\right]={\Lambda\over6}
 \ .\label{C3.24k}
\end{equation}
Eqs. (\ref{C3.24ab}) imply $\Omega_{,v}=A(u,v)(\xi+\bar\xi)$.
Eq. (\ref{C3.24c}) gives $A_{,v}=0$ so that $\Omega=A(u)(\xi+\bar\xi)v
+B(\xi,\bar\xi,u)$. By virtue of Eq. (\ref{C3.24de}) it must be of the form
$\Omega=A(u)(\xi+\bar\xi)v+C(u)\xi\bar\xi+D_1(u)\xi+D_2(u)\bar\xi+E(u)$.
Eq.  (\ref{C3.24f}) combined with Eq. (\ref{C3.24g}) gives
$dA/du=D_1+D_2$ so that $\Omega=A(u)(\xi+\bar\xi)v+C(u)\xi\bar\xi
+\pul dA/du(\xi+\bar\xi)+D(u)(\xi-\bar\xi)+E(u)$
with $A,C,E$ being real functions of $u$, and $D(u)$ being pure imaginary.
We distinguish the possibilities:

\centerline { 1) $A=0$}

\noindent
In this case Eq. (\ref{C3.24hi}) implies
$\Omega=C\xi\bar\xi+D(\xi-\bar\xi)+E$, with $C,E$  real constants,
$D$ a pure imaginary constant. Let (i) $C\not=0$.
Without loss of generality we can set $D=0$ by transformation
$\xi\rightarrow\xi'=\xi-D/C$.
If $E\not=0$, we can set $E=1$ by $\xi'=\xi/E$ which implies
$\Omega=C\xi\bar\xi+1$. Eq. (\ref{C3.24k}) gives $C=\Lambda/6$ so that
$\Omega=1+(\Lambda/6)\xi\bar\xi$, and the metric $\hat g_{\alpha\beta}$
takes the canonical form (\ref{E3.16}) of the $KN(\Lambda)I$ metric.
If $E=0$ we can make transformation $\xi'=1/(C\xi)$, after which
$\hat g_{\alpha\beta}$ takes the canonical form (\ref{E3.15}) of the
$KN$ metric.
Now (ii) $C=0$.
If $D\not=0$ we can set $E=0$ by $\xi'=\xi+E/(2D)$.
The relation (\ref{C3.24k}) gives $\Omega=\sqrt{\Lambda/6}(\xi-\bar\xi)$,
with $\Lambda$ necessarily being negative. It represents just another
coordinate form of the $KN(\Lambda)I$ metric since
by $\xi=(\xi'+i\sqrt{-{6/\Lambda}})/
(i\xi'+\sqrt{-{6/\Lambda}})$
we get the $KN(\Lambda^-)I$ metric in the canonical form (\ref{E3.16}).
If $D=0$ then  $\Omega=E$ so that $\hat g_{\alpha\beta}$ is the $KN$ metric
by transformation $\xi'=|E|\xi$.

\centerline { 2) $A\not=0$}

\noindent
Without loss of generality we can assume $A=1$ by using transformation
$w=A(u)v+\pul dA/ du$, $t=\int A^{-1}(u)\,du$. Therefore,
$\Omega=(\xi+\bar\xi)v+C(u)\xi\bar\xi+D(u)(\xi-\bar\xi)+E(u)$.
In this case Eq. (\ref{C3.24hi}) implies $H=2[{dC\over du}\xi\bar\xi+
\alpha(u)(\xi+\bar\xi)+{dD\over du}(\xi-\bar\xi)+{dE\over du}]$,
with $\alpha$ being an arbitrary real function of $u$. Eq. (\ref{C3.24k})
reduces to $\alpha(u)=C(u)E(u)+D(u)^2-\Lambda/6$, whereas
Eq. (\ref{C3.24j}) is satisfied identically. The solution given by
$\hat g_{\alpha\beta}$ is then conformally flat since
$g_{\alpha\beta}$ for $H$ of this form is conformally flat
( $C_{\alpha\beta\gamma\delta}=0$). Therefore, the solution describes
 Minkowski, de Sitter or  anti-de Sitter spacetime, according to
the sign of $\Lambda$. It is interesting to
notice that although $\hat g_{\alpha\beta}$ of this form describes
a conformally flat vacuum solution, the $KN$ solution to which it is conformal
(given by $g_{\alpha\beta}$ with $H$ of the same form) is conformally
flat but {\it not} necessarily a vacuum solution. In general, it is
a pure radiation solution, becoming a vacuum solution (Minkowski)
only for $C=$const. In particular, if $\alpha=0$ and $D,E=$const., then
the conformally flat $KN$ pure radiation solution given by
$H=2{dC\over du}\xi\bar\xi$ is  Wils' solution (3.10) \cite{Wils} for $N=0$
(where $Q+\bar Q={dC\over du}$, $v\rightarrow{v/{(\xi+\bar\xi)^2}}$).
The solution was used by Wils as an explicit counterexample
of theorem 32.17 in \cite{KSMH} according to which there are
no other conformally flat pure radiation solutions besides the special $PP$
wave of McLenaghan {\it et al.} \cite{McL}.

We have thus shown that all vacuum solutions conformal
to the $KN$ class are
1)  $KN(\Lambda)I$ solutions,
2)  $KN$ solutions themselves ($KN=KN(\Lambda=0)I$),
3)  Minkowski, de Sitter and anti-de Sitter space-times
 (these are special cases of  $KN(\Lambda)I$ given by
  $KN(\Lambda)I[H=H_c]$).
Therefore, {\it all vacuum solutions conformal to the $KN$ class belong to the
$KN(\Lambda)I$ class}, which proves Proposition 2 in Section II.

\end{document}